\documentclass[prb,twocolumn,showpacs,showkeys,floatfix]{revtex4}
\usepackage{amsmath,amsfonts,amsthm,graphicx}
\usepackage{graphicx}
\usepackage{graphics}
\usepackage{dcolumn}
\usepackage{bm}

\newcommand{\beq}{\begin{equation}}
\newcommand{\eeq}{\end{equation}}
\newcommand{\beqnar}{\begin{eqnarray}}
\newcommand{\eeqnar}{\end{eqnarray}}
\newcommand{\bfig}{\begin{figure}}
\newcommand{\efig}{\end{figure}}

\begin{document}
\title{Topological properties and edge states in a driven modified dimerized chain}

\author{Fatemeh Askari Shahid, Hosein Cheraghchi}
\email{cheraghchi@du.ac.ir} \affiliation{School of Physics,
Damghan University, 36716-41167, Damghan, Iran}
\date{\today}
\newbox\absbox

\begin{abstract}
We investigate topological phases induced by a driven electric field coupled to a dimer chain (a model for polyacetylene) at high frequency regime. It is shown how the topological invariant of the system can be controlled by the field amplitude. Furthermore, in the presence of a time-periodic electric field, the effect of the next-nearest neighbour hopping amplitudes on topological properties is studied. Breaking of the inversion symmetry causes to remove the degeneracy of zero edge states. The fractional Zak phase which is now measurable by ultra-cold atoms in one dimensional optical lattice is also calculated. For calculating modified band structure, we also develop a general Floquet-Bloch approach for systems under application of a potential with lattice and time translation invariance.
\end{abstract}

\pacs{}

\keywords{Topological invariant, SSH model, Zak phase} \maketitle

\section{Introduction}
Additional to the topological insulators originating from spin-orbit interaction\cite{HgTe,Hasan}, recently, the attention has been paid to inducing non-trivial topological phases by application of time-periodic perturbations\cite{photovoltaic,kitagawa2010,kitagawa2011}. Those topological phases which are sometimes disappeared in the undriven trivial systems\cite{Foa2015}. By using recent advances in experiment, engineering of quantum properties is manifested such that a topological band structure is achievable to create by application of an external irradiation\cite{FB_TI,PhotonicFTI}. Furthermore, a quantum Hall state can be realized in a time-periodic perturbation\cite{photovoltaic,Auerbach}. 

Motivated by the new possibility in the measurement of topological invariants\cite{direct_measure,Haldane_exp}, one dimensional systems with non-trivial properties have recently attracted much attention to study\cite{Foa2015,SSH_NNN,Hidden_symmetry}. Ultra-cold atoms trapped by optical lattices have prepared an appropriate substrate for realizing topological invariants and measuring the Zak phase as the parameter for characterizing topological properties of 1D systems\cite{direct_measure}. The simplest one dimensional model which exhibits topological band structure is so called Su-Schrieffer-Heeger (SSH) model which belongs to the BDI symmetry class; a model for describing polyacetylene\cite{SSH}. This model can be mapped into many other systems such as; graphene nanoribbons\cite{Zak}, sp-orbital optical ladder systems\cite{sp_ladder}, off-diagonal bichromatic optical lattices\cite{AAH}. Such a dimerized chain has a topological non-trivial phase depending on the ratio of the intera-cell to inter-cell hopping parameters\cite{Hatsuqai}. However, illumination of an $ac$ electric field can displace the boundary of trivial to non-trivial phases and induce some extra edge states which are controlled by the renormalized parameters arising from $ac$ field\cite{platero}. On the other hand, the extended SSH system with considering next-nearest neighbour hopping amplitude shows Haldane's phase diagram when hopping amplitudes are cyclically modulated by an additional parameter\cite{SSH_NNN}. However, for manifesting such a system  

In this work, for calculating the band structure of a quantum lattice in the presence of time-periodic perturbation, we present a general and straightforward approach based on the Floquet-Bloch (FB) states for a lattice with basis in its unit cell. In different driving regimes, photon-assisted quasi-energy spectrum and F‌‌B states are derived for a modified driven SSH system in the presence of next-nearest neighbour (NNN) couplings. The NNN couplings break the inversion symmetry which results in a break at the zero mode's degeneracy. At high frequency regime, the quasi-energy spectrum and its topological invariant (the Zak phase) are investigated to characterize the edge state dependency on the $ac$ field parameters. Motivated by the direct measurement of the Zak phase in onsite-staggered ultra-cold atoms trapped in 1D optical lattice \cite{direct_measure}, the Zak phase variation inducing by the NNN hopping parameter is studied.   

This paper is organized as the following: In section II, the FB band theory is presented in a general form. As an application, we introduce modified driven SSH Hamiltonian in section III. The FB band structure for different cases will be presented in section IV for high frequency regime. The next section is addressed to the Zak phase calculation and the edge states tracing. Conclusion section summarizes our results.      
\section{Floquet-Bloch band theory on a crystal with basis}
In a periodically driven quantum lattice, Hamiltonian is periodic
in time and space
$H(\vec{r},t)=H(\vec{r}+\vec{R},t)=H(\vec{r},t+T)$, where
$\vec{R}$ and $T=2 \pi /\omega$ are the lattice vector and
period of time. The time-dependent Schrodinger equation is as the following form.
\begin{align}
\mathcal{H}_{F}(t)\Psi(\vec{r},t)=0 \label{Schrodinger}
\end{align}
where Floquet Hamiltonian is defined as $\mathcal{H}_{F}(t) =  H(t) -i \dfrac{\partial}{\partial t}$.
Here $H(t)$ is the time-periodic Hamiltonian of a lattice and hence wavefunction $\Psi$ must be a Bloch type both in time and space. As a result, the wavefunction in a one period evolution changes by a pure phase factor leading to invariance of the observable value  $<\Psi|A|\Psi>$. Therefore, we have $\Psi(\vec{r},t)=e^{-i\epsilon t} \psi(\vec{r},t)$, where $\psi$ as a FB state is a periodic function of time and obeys the Bloch theorem in space as well. Here $\epsilon$ is called the quasi-energy of FB states. Replacing this type of wavefunction in Eq.\ref{Schrodinger} transforms Schrodinger equation to an eigenvalue equation form.
\begin{equation}
H_F(t)\psi(\vec{r},t)=\epsilon \psi(\vec{r},t)
\end{equation}
Remembering the tight-binding method, we can in principle construct the crystal states consisting of a linear combination of localized atomic orbitals \cite{grosso}. Therefore, one can correspondingly construct the FB sum of $\vec{k}$ vector as the following:\\
\begin{equation}
\psi _{ \lambda, \alpha, n }(\vec{k},\vec{r},t)
=N^{-\frac{D}{2}} 
\sum_{\vec{R}} e^{i\vec{k}.\vec{R} - in\omega t} \phi_{\alpha ,n}(\vec{r}-\vec{R}-\vec{\tau}_{\lambda})  \label{FB_state}
\end{equation}
where $\phi_{\alpha ,n}(\vec{r}-\vec{R}-\vec{\tau}_{\lambda})$ is the
$\alpha$ 'th Floquet atomic orbital centered in the reference unit cell
for the atom in a position $\vec{\tau}_{\lambda}$ in the unit cell (for $\lambda$ 'th basis in the unit cell). Here $n$, $D$ and $N$ are the quantum
number characterizing Floquet bands, the system dimension and the number of unit cell in the crystal, respectively. $|\phi_{\vec{R},
\vec{\tau}, \alpha ,n }>$ is a periodic state in space. The above FB sum preserves both the Bloch theorem and time
invariant of the FB state.
\begin{align}
T_{\vec{R}^{\prime}} |\psi _{\lambda \alpha,n} (\vec{k}, \vec{r}, t) >&=e^{i\vec{k}.\vec{R}^{\prime}}  |\psi _{\lambda \alpha,n} (\vec{k}, \vec{r}, t) >, \\
T_{\mathcal{T}} |\psi _{\lambda \alpha,n} (\vec{k}, \vec{r}, t)>&=  |\psi _{\lambda \alpha,n} (\vec{k}, \vec{r}, t)>
\end{align}
where $T_{\vec{R}^{\prime}}$ and $T_{\mathcal{T}}$ are
translational operators in space and time. The FB sum defined in Eq.(\ref{FB_state}) is used as the basis function for constructing a general solution for the crystal wave function of $\vec{k}$ in the form 
\begin{align}
\Phi(\vec{k},\vec{r},t)=\sum_{\lambda,\alpha,n} C_{\lambda,\alpha,n}(\vec{k})\psi _{ \lambda, \alpha, n }(\vec{k},\vec{r},t).
\end{align}
where the coefficients $C_{\lambda,\alpha,n}(\vec{k})$ are determined by variational method. By using the above expansion of crystal wave function on FB sums, one can calculate quasi-energies and eigenfunctions of Floquet Hamiltonian by application of the variational principle. The quasi-energy spectrum and crystal eigenfunctions are obtained from the following determinant. 
\begin{equation}
||[ \mathcal{H}_{F}(\vec{k})]_{\lambda \alpha ,\lambda ^{\prime} \alpha ^{\prime} }^{n,m} -\epsilon \delta_{n,m} \delta_{\lambda \alpha,\lambda^{\prime} \alpha^{\prime}}||=0\nonumber 
\end{equation}
The matrix elements of Floquet Hamiltonian in the basis of FB sums (Eq.(\ref{FB_state})) are written in the following form.
\begin{eqnarray}
\lefteqn{[ \mathcal{H}_{F}(\vec{k})]_{\lambda \alpha ,\lambda ^{\prime} \alpha ^{\prime} }^{n,m} }  \\ 
& = &\ll \psi _{\lambda,\alpha, n}(\vec{k}, \vec{r},t)| \mathcal{H}_{F}(t) |\psi _{\lambda ^{\prime} ,\alpha ^{\prime}, m }(\vec{k},\vec{r},t)\gg   \nonumber \\ 
&=&  {G}^{n,m}_{\lambda \alpha ,\lambda ^{\prime} \alpha ^{\prime} }(\vec{k})-  n\omega \delta_{n,m}\delta_{\lambda,\lambda^{\prime}}\delta_{\alpha ,\alpha^{\prime}}   \nonumber  \label{FB_hamiltonian}
\end{eqnarray}
where
\begin{eqnarray}
{G}^{n,m}_{\lambda \alpha ,\lambda ^{\prime} \alpha ^{\prime} }(\vec{k})&=&\sum_{\vec{R}^{\prime\prime} }e^{-i\vec{k}.\vec{R}^{\prime\prime} } \left[ \widetilde{H^{m-n}} \right]_{\vec{R}^{\prime\prime} ,\vec{\tau_{\lambda}},\alpha;0,\vec{\tau}^{\prime}_{\lambda^{\prime}},\alpha^{\prime}} \nonumber \\
\end{eqnarray}
In the above calculation, the time averaging is denoted to $\ll..\gg=1/T \int_{0}^{T} dt <..>$. Furthermore, we use orthogonality condition of the Floquet atomic orbitals as the following $ \ll \phi_{\vec{R}, \vec{\tau}, \alpha ,n}|\phi_{\vec{R}^{\prime},\vec{\tau}^{\prime},\alpha^{\prime} ,m }\gg= \delta_{n,m}\delta_{\vec{R},\vec{R}^{\prime}}\delta_{\vec{\tau}, \vec{\tau}^{\prime}}\delta_{\alpha ,\alpha^{\prime}}$ . Here, Fourier transformation of the time-periodic Hamiltonian is defined as $ \widetilde{H^{m-n}}=\dfrac{1}{T} \int_{0}^{T} H(t)e^{i(m-n) \omega t} dt$. The expectation value of $\widetilde{H}$ on the Floquet atomic orbitals is written as 
\begin{align}
\left (\widetilde{H^{m-n}}\right)_{\vec{R}^{\prime\prime} ,\vec{\tau},\alpha;0,\vec{\tau}^{\prime},\alpha^{\prime}} = <\phi_{\vec{R}^{\prime\prime}, \vec{\tau}, \alpha ,n}| \widetilde{H^{m-n}}|\phi_{0,\vec{\tau}^{\prime},\alpha^{\prime},m }>
\end{align}
where $"0"$ denotes to the reference unit cell. Consequently, to have the spectrum of driven systems it is enough to calculate matrix elements of FB Hamiltonian $G$ as represented in Eq.\ref{FB_hamiltonian}. Diagonalizing the Floquet Hamiltonain results in the quasi-energies of driven system. 

\begin{figure}
\centering
\includegraphics[scale=0.27,angle=270]{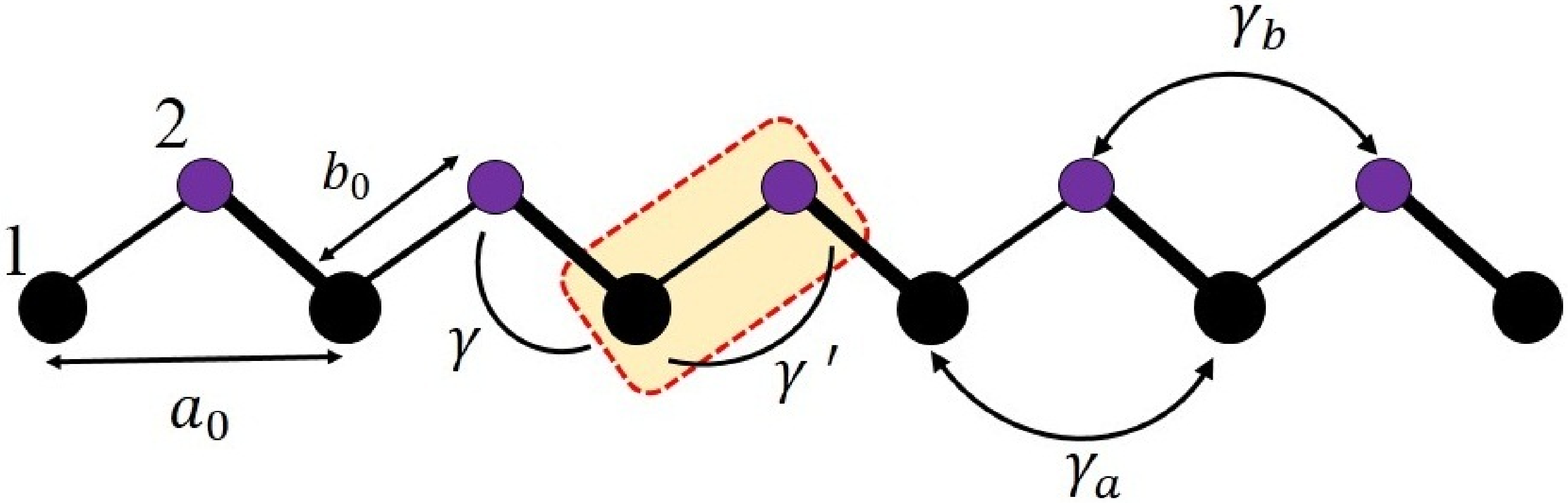}
\caption{A schematic view of driven modified SSH model where intera-cell and inter-cell hoppings are $\gamma^{\prime}$ and $\gamma$, respectively. Hoppings between atoms in $A$ and $B$ sublattices are $\gamma_a$ and $\gamma_b$, respectively.}
\label{ssh_fig1}
\end{figure}

\section{Driven modified SSH model}
A well-known simplest model for describing topological insulator properties is SSH model. This model is a spinless fermion hopping in one-dimensional lattice with staggered hopping amplitudes. In the Driven modified SSH Hamiltonian, nearest-neighbour(NN) and next-nearest-neighbour(NNN) hoppings are both time dependent inducing by an $\it{ac}$ electric field.
 
\begin{eqnarray}
H(t) &= & \sum_i [(\gamma^{\prime}(t) c^{\dagger}_{a,i} c_{b,i}+\gamma(t) c^{\dagger}_{b,i-1} c_{a,i}+ C. C.) \nonumber \\ &+& (\gamma_a(t) c^{\dagger}_{a,i-1} c_{a,i}+ \gamma_b(t) c^{\dagger}_{b,i-1} c_{b,i})+C. C.) ]
\end{eqnarray}
where hopping parameters depend on time as the following $\gamma_{i,j}(t)=\gamma^0_{i,j} \exp[\frac{2 \pi}{\Phi_0} \int_{\vec{r}_{j}}^{\vec{r}_{i}} \vec{A}(t).d\vec{l}]$. $\Phi_0$ is the quantum of magnetic flux. The vector potential is originated from a time-periodic electric field written in a Weyl gauge $\vec{E}=-\partial\vec{A}/\partial t$. Here we neglect magnetic component of illuminated laser. The vector potential results in a phase shift in the hopping amplitude connecting site $i$ to $j$. As depicted in Fig.\ref{ssh_fig1}, hoppings at zero field, $\gamma^0_{ij}$, are equal to $\gamma^{\prime}$ for intra-cell hopping, $\gamma$ for inter-cell hoppings, $\gamma_a$ for hopping between atoms in $A$ sublattice, and $\gamma_b$ for hopping between $B$ atoms. The vector potential for monochromatic waves with linear polarization is defined as $A(t)=A_0 \sin(\omega t)$, where $A_0=E_0/\omega$ is related to the amplitude of the electric field.

FB Hamiltonian can be derived using formula \ref{FB_hamiltonian} for arbitrary $n, m$. In the SSH model, the number of basis in the unit cell and number of orbitals on each site are as $\lambda=2, \alpha=1$, respectively. 

\begin{figure}
\centering
\includegraphics[scale=0.55]{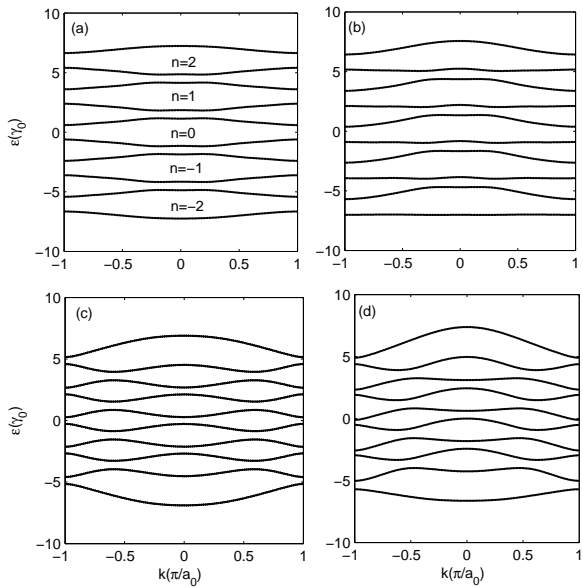}
\caption{ The quasi-energy spectrum of the driven modified SSH model for different cases at high frequency regime $\omega=3$: a) considering only nearest neighbour (NN) hoppings with $\gamma_a=\gamma_b=0$, b) by adding next nearest neighbour (NNN) hoppings to the Hamiltonian $\gamma_{a}=0.2,\gamma_{b}=0.1$. Here $\lambda=\gamma^{\prime}/\gamma$ is lesser than the unity ($\lambda=0.3$). The quasi-energy spectrum in a frequencies comparable with the NN hoppings $\omega=1.2$ and $\lambda=1.2$ for the two cases: c) NN approximation, d) in the presence of NNN hoppings with $\gamma_{a}=0.3,\gamma_{b}=0.1$. In all cases, the amplitude of the ac electric field is considered to be $E_{0}=0.1$. All energies are scaled to $\gamma=1$.}\label{quasi-energy}
\end{figure}

\begin{eqnarray}
{G}_{n,m}(k) = 
\begin{pmatrix}
2\gamma_{a} \cos (ka_{0}) J_{m-n}(z)&  \rho(k) \\
\\
\rho^* (k) & 2\gamma_{b} \cos (ka_{0}) J_{m-n}(z)
\end{pmatrix} \label{SSH_Hamiltonian}
\end{eqnarray}
where
\begin{align}
\rho (k) &= \gamma^{ \prime}  J_{n-m}(x) + \gamma J_{m-n}(y) e^{ika_{0}}\nonumber \\
\end{align}
$$ x=\dfrac{2 \pi A_{0}b_{0}}{\Phi_{0}}\quad,y=\dfrac{2 \pi A_{0}(a_{0} - b_{0})}{\Phi_{0}}\quad,z=\dfrac{2 \pi A_{0}a_{0}}{\Phi_{0}}
$$
In the above equation we have used the following definition of the Bessel function $
J_{p} (x) = \dfrac{1}{T}\int^{T} _{0} e^{i x \sin (\omega t) - ip \omega t} dt$. Let us first look at the quasi-energy spectrum of the bulk system for $n,m=2$ in the two regimes of high and low driving frequency. Fig.\ref{quasi-energy} (a,b) represents Floquet side bands separated by the label $n$ in the high frequency regime. Different replicas which are called Floquet side bands are related to $n$ number of photons assisted in the spectrum. In the degenerated energies of spectrum, the driving induced gaps appear around the energy $\hbar \omega /2$. It should be noted that these dynamical gaps are not originated from breaking of time reversal symmetry \cite{Foa2015}. These dynamical gaps are proportional to the amplitude of the vector potential.  

\begin{figure}
\centering
\includegraphics[width=8cm,height=6cm]{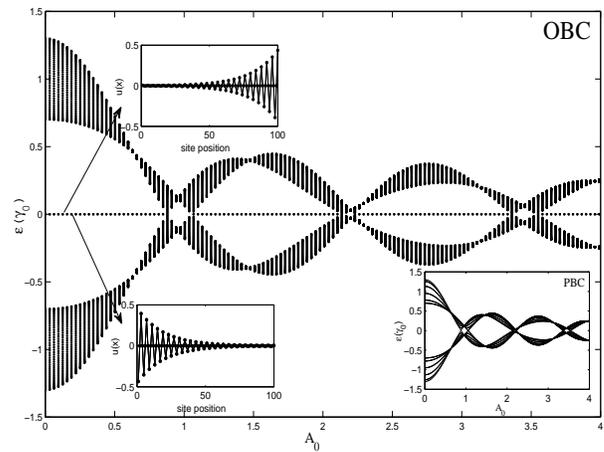}
\caption{Quasi-energy spectrum of the driven modified SSH model as a function of the amplitude of the vector potential $A_0$ at high frequency regime $\omega =10$ and in OBC with 50 dimers in the chain. Here only NN approximation is considered $\gamma_{a}=\gamma_{b}=0 $. In this case we consider $\lambda =0.3$. Left inset figures indicate eigen wave function in terms of site positions. In the high frequency regime, right-down inset figure shows quasi-energy band structure of the driven modified SSH model for different values of the wave vectors in PBC.} \label{NN_spectrum}
\end{figure}
In high frequency regime ($\hbar \omega \gg \gamma^0$), Floquet bands are decoupled from each other and so FB Hamiltonian in Eq. \ref{SSH_Hamiltonian} becomes block diagonal. High frequency regime corresponds to $x,y,z \rightarrow 0$ and so all Bessel functions goes to zero for $n\neq m$. In this case, Hamiltonian \ref{SSH_Hamiltonian} can be represented in terms of the Pauli matrices.
\begin{align}
{G}_{n}(k) = E_I(k)I+\vec{d}(k).\vec{\sigma}
\end{align}
where $I$ and $\vec{\sigma}=(\sigma_x,\sigma_y,\sigma_z)$ are referred to the identity matrix and Pauli matrices, respectively. The components of vector $\vec{d}$ and diagonal terms $E_I(k)$ are 
\[
  \begin{cases}
   E_I(k)=(\gamma_a+\gamma_b)J_{0}(z)\cos(ka_0)\nonumber \\
   d_x(k)=\gamma^{\prime} J_{0}(x)+\gamma J_{0}(y) \cos(ka_0) &\\
   d_y(k)=\gamma J_{0}(y) \sin(ka_0)  & \\
   d_z(k)= (\gamma_a-\gamma_b)J_{0}(z) \cos(ka_0)&\\
  \end{cases}
\].

Here if $\gamma_a=\gamma_b$, then $d_z=0$ and therefore the inversion symmetry $\sigma_x G(k) \sigma_x=G(-k)$ is preserved. As a result, there is no gap opening. However, in this case, the energy shift is not zero $E_I(K) \neq 0$ which leads to breaking of the particle-hole symmetry and also chiral symmetry. However, the energy shift is zero when $\gamma_a=-\gamma_b$, while $d_z \neq 0$. Therefore, in this case, the inversion symmetry is broken while the particle-hole symmetry is preserved. So, a band gap is opened in the spectrum. In the special amplitude of the electric field where $J_0(z)=0$ at $z=z*$ as the zeros of the Bessel function, the chiral symmetry $\sigma_z G(K) \sigma_z=-G(k)$ and also the inversion symmetry are both preserved leading to degenerate zero-mode edge state.  
\begin{figure}
\centering
\includegraphics[width=8cm,height=6cm]{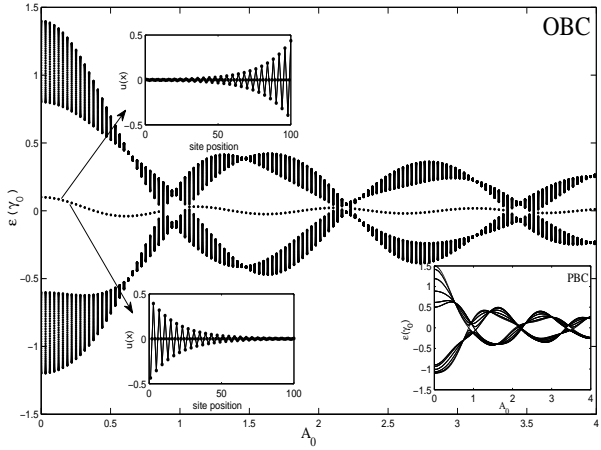}
\caption{Quasi-energy spectrum of the driven modified SSH model as a function of $A_0$. Here $\gamma_{a}=\gamma_{b}=0.1$ and $\lambda=0.3$. Left inset figures indicate decaying of the two degenerate edge modes. Band structure of the bulk system (PBC) is presented in the right-down inset figure for different wave vectors.\label{symmetric}}
\end{figure}

In high frequency regime and for $n=m$, the quasi-energy of Hamiltonian shown in Eq.\ref{SSH_Hamiltonian} is read as $\epsilon_{\pm}(k)=E_I(k)\pm |\vec{d}|$, where the condition for the gap closing point at $k=\pi/a_0$ is derived as the following.
 $$[\gamma^{\prime}J_0(x)-\gamma J_0(y)]^2+[\gamma_a-\gamma_b]^2J_0(z)^2=0$$
In the case $\gamma_a=\gamma_b$, the gap closing point would occur at $\Lambda=\frac{\gamma^{\prime}}{\gamma}\frac{J_0(x)}{J_0(y)} =1$ and a topological phase is expected to appear for $\Lambda<1$. In general case of $\gamma_a \neq \gamma_b$, the bulk system has a gap with no electron-hole symmetry in the band structure.  
\section{Band Structure in presence of driven electric field}
At high frequency regime, it will be shown that based on the value of $\Lambda$, there is a trivial to non-trivial phase transition. In fact, by variation in the amplitude of the vector potential ($A_0$), one can control topological properties of the system. To show the edge modes, the spectra of Floquet quasi-energy in terms of $A_0$ is investigated for both: the limited number of dimer chains (50 dimers) in the open boundary condition (OBC) and also the periodic boundary condition (PBC). In all following cases, driving frequency is considered to be $\omega=10$.

{\it Case 1}: In the NN approximation, there are two zero edge state modes with double degeneracy which are seen in Fig.\ref{NN_spectrum}. In this case, because $\gamma_a=\gamma_b=0$, the edge modes are protected by both the electron-hole and inversion symmetries. There is no gap opening in PBC indicating in the right-down inset figure \ref{NN_spectrum} and the band structure is symmetric around the zero mode. The left inset figures \ref{NN_spectrum} show that paired edge states located on the opposite edges of dimer chain are pinned to the zero energy and they decay exponentially with the localization length $\xi$ that depends on $\Lambda$ as the following form, $\Lambda \approx \exp [-a_0/\xi]$. So the localization length enhances when $\Lambda \rightarrow \Lambda_{cr.}$, where $\Lambda_{cr.}=1-\frac{1}{M+1}$. Here $M$ is the number of dimers in a finite dimer chain. There is no edge states for the amplitudes of the electric field satisfied $\Lambda > \Lambda_{cr.}$. In fact, by application of the driven field, hopping energies are renormalized by zero Bessel functions. So hopping energies are smaller than the hopping parameters of the undriven SSH model. Therefore, the band width of the driven SSH model is thinner than the undriven one.  
\begin{figure}
\centering
\includegraphics[width=8cm,height=6cm]{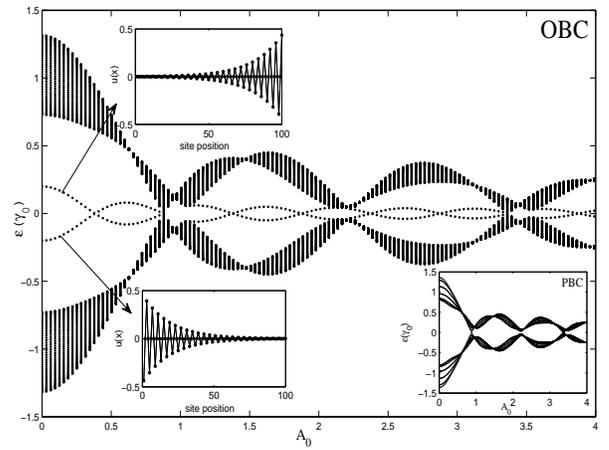}
\caption{Quasi-energy spectrum of the driven modified SSH model as a function of $A_0$. Here $\gamma_{a}=-\gamma_{b}=0.2$ and $\lambda=0.3$. Left inset figures indicate decaying of the two degenerate edge modes. Band structure of the bulk system (PBC) is presented in the right-down inset figure for different wave vectors.}\label{antisymmetric}
\end{figure}

{\it Case 2}: Let us consider contribution of the NNN hoppings in the Hamiltonian. At the first step, we investigate the symmetric NNN hoppings in which $\gamma_a=\gamma_b \neq 0$. Quasi-energy spectrum of Fig.\ref{symmetric} shows breaking of the electron-hole symmetry in the spectrum. In addition, the edge modes have no zero energy. The edge modes cross zero energy only at the special electric fields as  $z=z^*$. In this case, the inversion symmetry is preserved and so there is no gap in the bulk system (PBC). As we will show later, adding the term $E_I(k)$ whenever $d_z(k)=0$ just shifts energy spectrum and does not change the topological phase diagram in compared to the {\it case} $1$. In this case, the chiral symmetry is broken.   

\begin{figure}
\centering
\includegraphics[width=8cm,height=6cm]{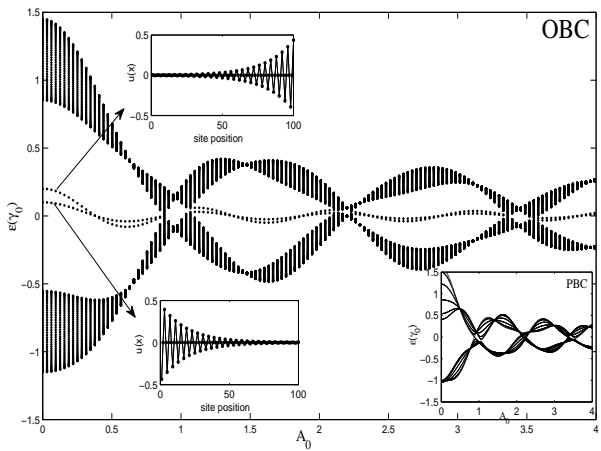}
\caption{Quasi-energy spectrum of the driven modified SSH model as a function of $A_0$. Here $\gamma_{a}=0.2, \gamma_b=0.1$ and $\lambda=0.3$. Left inset figures indicate decaying of the two degenerate edge modes. Band structure of the bulk system (PBC) is presented in the right-down inset figure for different wave vectors.}\label{generic_case}
\end{figure}

{\it Case 3}: It is interesting to have a look at the spectrum of the anti-symmetric NNN hoppings in Fig.\ref{antisymmetric} where $\gamma_a=-\gamma_b$. In this case, the degeneracy of the two edge modes located at the opposite ends of the chain is broken. The spectrum is symmetric around zero energy. Here $d_z \neq 0$, so a gap is opened in the bulk system (PBC) except for the points $z=z^*$. In addition, the Zak topological phase differs from the Zak phase related to the {\it cases} $1$ and $2$. Crossing points of the edge states (OBC) occur at the points $z=z^*$.

{\it Case 4}: In general case, NNN hoppings are considered to be $\gamma_a \neq \gamma_b$. In this case, there is no symmetry at all. So as shown in Fig.\ref{generic_case}, band structure is asymmetric and also gapped. Furthermore, the edge modes are non-degenerate and have non-zero energy. Emergence of the edge modes crossing each other at the points $z=z^*$, proposes that these systems are topologically non-trivial. All the above cases are calculated for $\lambda <1$, where in the undriven SSH model there exists a non-trivial topological phase. However, for $\lambda >1$ where the undriven system is in the trivial phase, $ac$ electric field can induce the non-trivial topological phases as seen in Fig.\ref{landaGT1}. 
\section{Zak phase for driven modified SSH model}
To elucidate the relation between the existence of the edge modes in a finite chain with the topological invariants in the bulk system, the Zak phase is calculated based on the Hamiltonian \ref{SSH_Hamiltonian} for the different cases. The Zak phase is a parameter for measuring the solid angle that the pseudo-spinor of $G_0(k)$ rotates on the Bloch sphere when $k$ spans the Brillouine zone. The pseudo-spinor of $G_0(k)$ can be written as:
\begin{eqnarray}
| \psi_+(k) \rangle = 
\begin{pmatrix}
\cos \frac{\theta(k)}{2} e^{-i\phi(k)} \\
\\
\sin \frac{\theta(k)}{2} 
\end{pmatrix} \nonumber \\ \\
|\psi_{-}(k) \rangle= \nonumber 
\begin{pmatrix}
\sin \frac{\theta(k)}{2} e^{-i\phi(k)}\\
\\
-\cos \frac{\theta(k)}{2}   
\end{pmatrix} \label{pseudo-spinor}
\end{eqnarray}

\begin{figure}
\centering
\includegraphics[width=8cm,height=6cm]{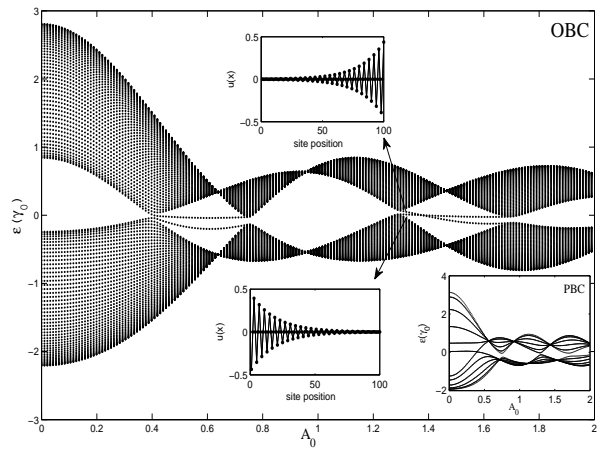}
\caption{Quasi-energy spectrum of the driven modified SSH model as a function of $A_0$. Here $\gamma_{a}=0.5, \gamma_b=0.1$ and $\lambda=1.5$. Left inset figures indicate decaying of the two degenerate edge modes. Band structure of the bulk system (PBC) is presented in the right-down inset figure for different wave vectors.} \label{landaGT1}
\end{figure}
where $\phi(k)=\arctan(d_y/d_x)$ and $\theta(k)=\arccos(d_z/d)$. It should be noticed that the constant term $E_I(k)$ does not modify the pseudo-spinor. For the case of $d_z=0$, the pseudo-spinors evolve with the wave vector in the equatorial plane of the Bloch sphere when $\theta(k)=\pi/2$. In this case, the Zak phase is multiple of $\pi$. However, in the modified SSH model with NNN hoppings ($d_z \neq 0$), the pseudo-spinors move away from the equatorial plane leading to Zak phase value which is not necessarily a multiple of $\pi$. For a Bloch function $\psi(k)$, the Zak phase can be expressed as the following:
 $$\Phi_Z=i \int_{-\pi/a_0}^{\pi/a_0} \langle \psi(k) |\partial_k|\psi(k) \rangle dk$$ 
If one consider the NNN hoppings, the Zak phase is changed from $\Phi_{Z}(0)$ to $\Phi_{Z}(d_z)$ for the upper and lower bands.
\begin{equation}
\delta \Phi_{Z,\pm}=\Phi_{Z,\pm}(d_z)-\Phi_{Z}(0)=\mp\int_{-\pi/a_0}^{\pi/a_0}\frac{d_z}{2d}\partial_k \phi(k) dk \label{Zak_general}
\end{equation} 
where the Zak phase for the case of $d_z=0$ are written as the following form.
\begin{equation}
\Phi_{Z}(0)=\frac{1}{2}\int_{-\pi/a_0}^{\pi/a_0}\partial_k \phi(k) dk= \frac{1}{2}\int_{-\pi}^{\pi}\frac{1+\Lambda \cos(x)}{f_{\Lambda}(x)} dx \label{Zak_SSH}
\end{equation} 
where in these integrals, the azimuthal angle of the pseudo-spinor and the function $f$ are derived as $\tan \phi(k)=\frac{\sin (ka_0)}{\Lambda+\cos (ka_0)}$ and $f_{\Lambda}(x)=\sin^2 x+(\Lambda+\cos x)^2$. The integration result for $\Phi_{Z}(0)$ is $\Phi_Z=\pi$ if $\Lambda<1$, while $\Phi_Z=0$ if $\Lambda>1$. So there is a non-trivial topological phase which corresponds to the regions with the existence of the edge modes in OBC. Even if the undriven system is in the trivial phase in a regime with $\lambda >1$, it would be possible to tune the parameter $\Lambda$ by using applied $ac$ electric field such that the driven SSH model (for $d_z=0$) shows non-trivial topological phase (Fig.\ref{landaGT1}).
\begin{figure}
\centering
\includegraphics[width=8cm,height=6cm]{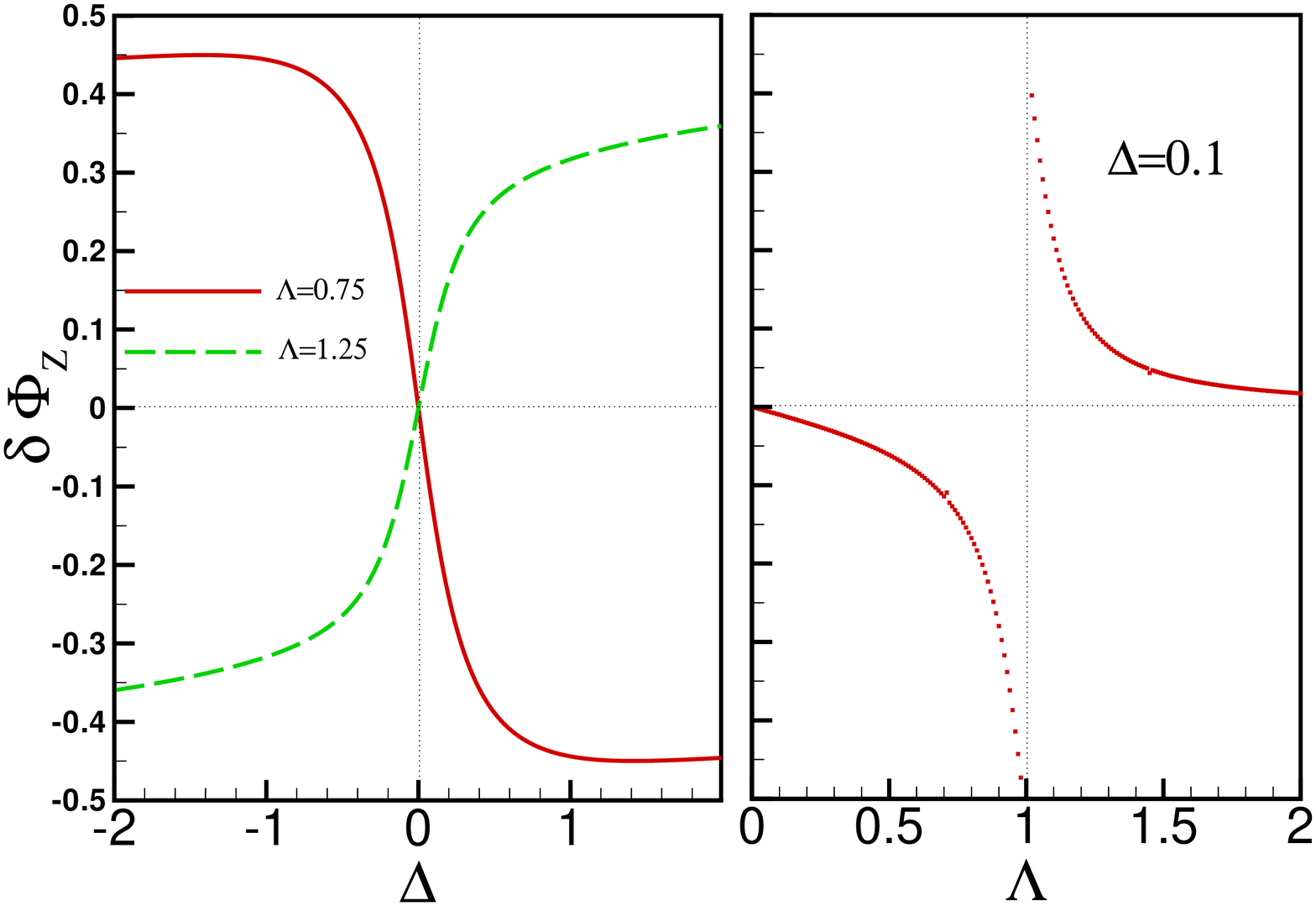}
\caption{The Zak phase difference $\delta \Phi_{Z,-}=\Phi_{Z,-}(\Delta)-\Phi_Z(0)$ in terms of a) $\Delta$ (for two regimes $\Lambda <1$ and $>1$) and b) $\Lambda$ (with $\Delta=0.1$). \label{Zak_dif} }
\end{figure}

\begin{figure}
\centering
\includegraphics[width=8.5cm,height=7cm]{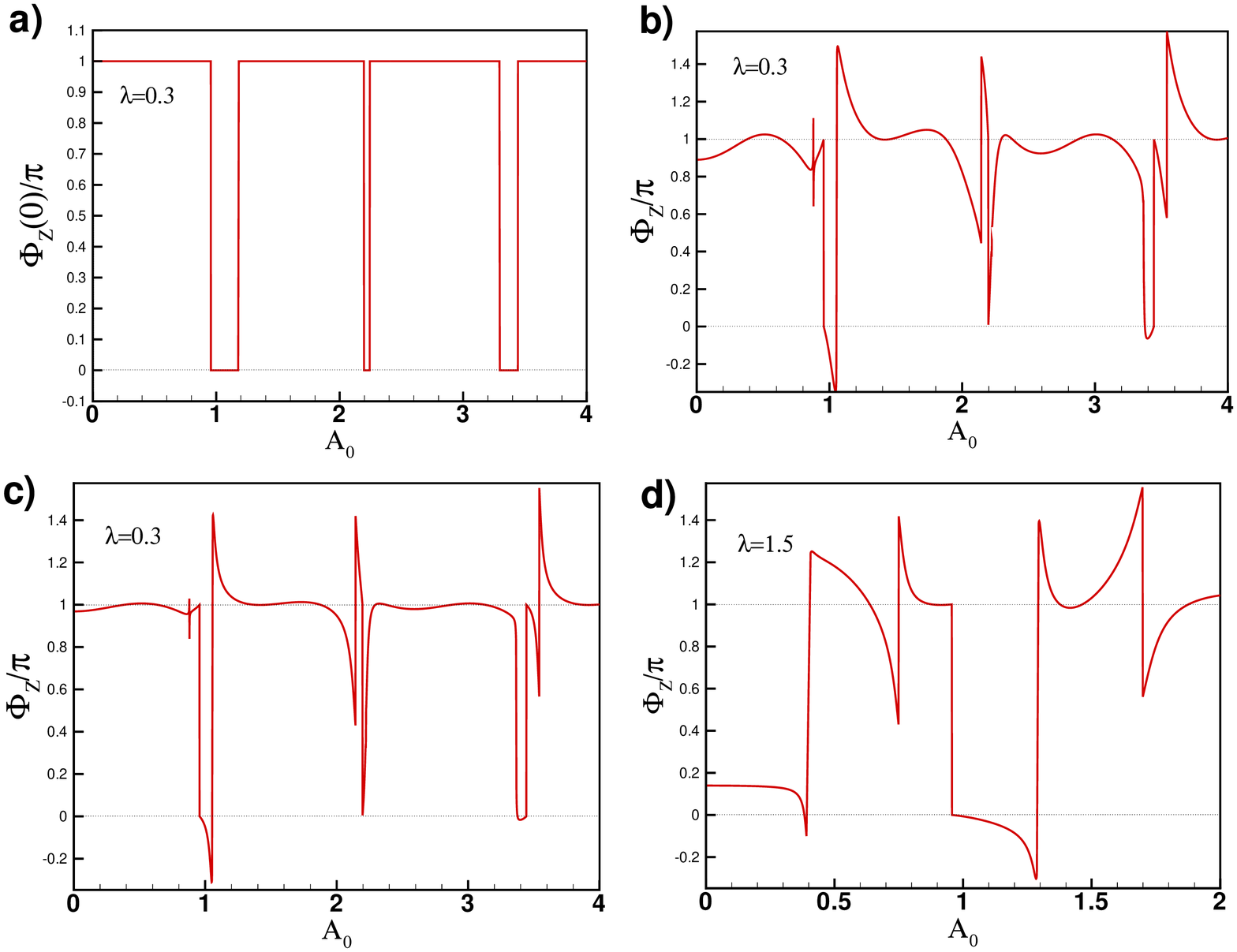}
\caption{The Zak phase for different cases as a function of the field amplitude. a) for the case with $t_a=t_b$ b) for $t_a=-t_b=0.2$ c) for $t_a=0.5, t_b=0.1$. All cases are in the regime  $ \lambda <1$. d) for $t_a=0.5, t_b=0.1$ with $\lambda=1.5$. Here the driven frequency is considered to be $\omega =10$.} \label{Zak_phase}
\end{figure}

The phase difference for the driven modified SSH model can be derived as the following.
\begin{equation}
\delta \Phi_{Z,\pm}=\mp \frac{1}{2}\int_{-\pi}^{\pi}\frac{(1+\Lambda \cos(x))\Delta \cos x}{f_\Lambda (x)\sqrt{f_\Lambda (x)+\Delta^2 \cos^2 x}}dx \label{Zak_eq} 
\end{equation} 
where the dimensionless parameter $\Delta$ is defined as $\Delta=\frac{(t_a-t_b)J_0(z)}{\gamma J_0(y)}$. The Zak phase in this case is a fraction of $\pi$. Considering the NNN hoppings $\Delta$ breaks the inversion and electron-hole symmetries.  Hence the edge modes are not protected and the non-trivial topological phase (for $\Lambda <1$) will be changed. Let us look at the results shown in Fig.(\ref{Zak_dif}.a) for the variation of the Zak phase difference for the lower band between the modified ($\Delta \neq 0$) and standard ($\Delta=0$) SSH model as a function of $\Delta$. By increasing $\Delta$, the pseudo-spin vector goes away from the equatorial plane leading to smaller solid angle variation whenever $k$ spans the first Brillouine zone. In the case $\Lambda >1$ where the phase of standard SSH model is trivial, by inducing NNN hoppings ($\Delta$), the Zak phase becomes non-zero, while for $\Lambda <1$, the Zak phase decreases to zero. Furthermore, as it is seen in Fig.(\ref{Zak_dif}.b), the deviation of the Zak phase from what we know from the standard SSH model, depends on $\Lambda$. In fact, the evolution of the pseudo-spin vectors with $k$ is large close to the transition point ($\Lambda=1$) . In other words, the variation in the solid angle inducing by $\Delta$ is large at the vicinity of the transition point. 

It should be noted that if we consider staggered on-site energy in the driven SSH model with the nearest neighbour approximation ($\gamma_a=\gamma_b=0$), again the z-component of the vector $\vec d$ is non-zero but independent of $k$ ($d_z=\Delta_0$). The formula derived in Eq.\ref{Zak_general} can be still used for calculating the Zak phase of staggered driven SSH model as well.

Finally, the Zak phase for the different cases was calculated by using equations \ref{Zak_SSH} and \ref{Zak_eq}. As seen in Fig.(\ref{Zak_phase}.a), the Zak phase profile for the case of the driven SSH model (case 1) and also the case with the symmetric NNN hoppings (case 2) are exactly similar. The Zak phase is zero or $\pi$ depending on the value of $A_0$. The regions with the $\pi$ value for the Zak phase (which correspond to $\Lambda <1$) are in agreement with the regions exhibiting the edge modes in figures \ref{NN_spectrum} and \ref{symmetric}. In the case with the symmetric NNN hoppings, the edge modes are degenerated but with non-zero energy. However, there exist non-trivial topological regions with the Zak phase value of $\pi$.  

For the cases $3$ and $4$, the Zak phase as a function of the field amplitude is presented in figures \ref{Zak_phase}.b and \ref{Zak_phase}.c. The results show some variations in the Zak values in compared to the Zak phase of the cases $1$ and $2$. Although the regions with non-trivial topological phases still remain unchanged, the Zak phase has a relatively large shift close to the transition points in compared to the Zak phase for the cases 1 and 2. The Zak phase at the points with $z=z^*$ is what one expects from the standard driven SSH model.

At the end, the Zak phase is calculated for the case with $\lambda >1$ where the undriven SSH system is in the trivial phase. However, as shown in Fig.(\ref{Zak_phase}.d), the Zak phase becomes nonzero in some regions of the field amplitudes. As shown in Fig.\ref{landaGT1}, the regions with the Zak phases around $\pi$ represent non-zero edge modes leading to non-trivial topological states.     
  
\section{Conclusion}
As a conclusion, we investigate topological invariants and the edge modes of the modified driven SSH model. In this paper, it is shown how application of an $ac$ electric field can change topological properties of the system. Since the hopping parameters are renormalized by means of time-periodic field, the Zak phase is dependent on the amplitude and frequency of the $ac$ electric field. At high frequency regime, the effect of the NNN hopping parameters on the edge modes and the Zak phase are studied leading to a phase variation measurable directly in ultra-cold atoms trapped in optical lattices. Furthermore, we also develop a general Floquet-Bloch band theory for $D$ dimensional lattice with a basis in its unit cell. This formalism prepares a straightforward and simple form for the FB band structure calculation.   

\section{Acknowledgement}
H.C. thanks the International Center for Theoretical
Physics (ICTP) for their hospitality and support during a visit
in which part of this work was done. We would like to acknowledge the instructive comments of Fatemeh Adinehvand in the early stages of the work.

\end{document}